\pdfoutput=1
\documentclass[12pt]{article}

\usepackage{jcappub}
\usepackage{amsmath}
\usepackage{amssymb}
\usepackage{braket}
\usepackage{graphicx}
\usepackage{hyperref}
\usepackage{mathrsfs}
\usepackage{subcaption}
\usepackage{empheq}
\usepackage{appendix}
\usepackage{natbib}
\usepackage[dvipsnames]{xcolor}
\usepackage{float}
\usepackage[top=1.2in, bottom=0.6in, left=0.8in, right=0.8in, includefoot]{geometry}

\title{Palatini inflation in models with an $R^2$ term}

\author[a,b]{I. Antoniadis,}
\author[c]{A. Karam,}
\author[c]{A. Lykkas,}
\author[c,a]{K. Tamvakis}

\affiliation[a]{LPTHE, Sorbonne Universite, CNRS, 4 Place Jussieu, 75005 Paris, France}
\affiliation[b]{Albert Einstein Center, Institute of Theoretical Physics, University of Bern, Sidlerstrasse 5, CH-3012, Bern, Switzerland}
\affiliation[c]{Physics Department, University of Ioannina, GR--45110 Ioannina, Greece}

\emailAdd{antoniad@lpthe.jussieu.fr}
\emailAdd{alkaram@cc.uoi.gr}
\emailAdd{alykkas@cc.uoi.gr}
\emailAdd{tamvakis@uoi.gr}

\abstract{The Starobinsky model, considered in the framework of the Palatini formalism, in contrast to the metric formulation, does not provide us with a model for inflation, due to the absence of a propagating scalar degree of freedom that can play the role of the inflaton. In the present article we study the Palatini formulation of the Starobinsky model coupled, in general nonminimally, to scalar fields and analyze its inflationary behavior. We consider scalars, minimally or nonminimally coupled to the Starobinsky model, such as a quadratic model, the induced gravity model or the standard Higgs-like inflation model and analyze the corresponding modifications favorable to inflation. In addition we examine the case of a classically scale-invariant model driven by the Coleman-Weinberg mechanism. In the slow-roll approximation, we analyze the inflationary predictions of these models and compare them to the latest constraints from the Planck collaboration. In all cases, we find that the effect of the $R^2$ term is to lower the value of the tensor-to-scalar ratio.}

\begin{document}

\maketitle

\section{Introduction}
Cosmological inflation is at present an attractive framework to address a number of issues of Big Bang cosmology. Its most promising aspect is the treatment of primordial fluctuations resulting in the large scale structures and the anisotropy in the temperature of the cosmic microwave background (CMB) we observe today. Models of inflation require the presence of a scalar degree of freedom (inflaton), either as a fundamental scalar field or incorporated into gravity itself, in general as an effective scalar degree of freedom. The Starobinsky model, featuring next to the Einstein-Hilbert term of general relativity (GR) a quadratic scalar curvature term, is persistently in agreement with ongoing observations, in contrast to most single scalar field models which, if not in disagreement, give borderline values for the relevant parameters. 

Generalizations of GR in the form of $f(R)$ theories~\cite{DeFelice2010, Sotiriou2010, Capozziello2011, Clifton2012, Nojiri2017a} have attracted a lot of attention in recent years. Such theories would in principle be suitable for inflation since they contain the scalar degree of freedom that can play the role of the inflaton. The Starobinsky model is one particular case of these theories\footnote{See~\cite{Cardenas2003, Bruck2015, Calmet2016, Kaneda2016, Mori2017, Ema2017, Wang2017, Gorbunov2018, He2018, Pi2018} for some modifications of the Starobinsky model with the addition of an extra scalar field.}. Any theory with an action of the form $\int\!\mathrm{d}^4x\sqrt{-g}\,f(R)$ can be reformulated as a scalar-tensor theory of gravity rewritten in terms of a scalar field non-minimally coupled to the Ricci scalar as $\int\!\mathrm{d}^4x\sqrt{-g}\left\{f'(\phi)R-V(\phi)\right\}$, where the potential is $V(\phi)=\phi\,f'(\phi)-f(\phi)$. 

It has been known for some time that an alternative variational principle leading to the equations of motion of GR is the Palatini or first order formalism (see~\cite{Sotiriou2010} for a review and~\cite{Bauer2008, Borunda2008, Olmo2011, Bauer2011, Tamanini2011, Enqvist2012, Borowiec2012, Racioppi2017, Rasanen2017, Fu2017, Stachowski2017, Szydlowski2017, Tenkanen2017, Markkanen2018, Carrilho2018, Enckell2018, Bombacigno2018, Enckell2018a, Kozak2018, Jaerv2018, Wang2018, Wu2018, Szydlowski2018} for various applications) in which, in addition to the metric $g_{\mu\nu}$, the connection $\Gamma_{\mu\nu}^{\rho}$ is treated as an independent variable. This does not constitute an additional assumption about the nature of the theory, but a different parametrization of the gravitational degrees of freedom. Another important point is that since the action contains derivatives of the connection, there is no need for a York-Gibbons-Hawking surface term. Within GR the two formulations are entirely equivalent. However, in the presence of a non-minimal coupling, this is not the case. When a fundamental scalar $\phi$ is coupled to gravity through a non-minimal coupling $f(\phi)R$ the metric and the Palatini formulations lead to different results. 

Another notable difference is the fact that, in the framework of the Palatini formulation, any $f(R)$ theory is entirely equivalent to the Einstein theory. A quick way to see this is by performing a Weyl rescaling of the metric $\bar{g}^{\mu\nu}=f'(\phi){g}^{\mu\nu}$ that transforms the action into the Einstein frame. In the Palatini formalism $R_{\mu\nu}$, being a function of $\Gamma$, will be unaffected and no kinetic term will be generated for the auxiliary scalar $\phi$. Thus, in the Einstein frame, there is no propagating scalar field and the action is just the Einstein action $\int\!\mathrm{d}^4x\sqrt{-\bar{g}}\bar{R}$ plus a potential term $V(\phi)/(f'(\phi))^2$ of the non-propagating scalar. Such a theory would not be suitable to describe inflation, since it contains no dynamical degree of freedom to take up the role of the inflaton. Nevertheless, all of this does not apply to $f(R)$ theories coupled to matter.

In the following section we set up the theory of the Starobinsky model, written in terms of an auxiliary scalar field and coupled, in general non-minimally, to a fundamental scalar field with a general potential in the framework of the Palatini formalism. Going to the Einstein frame and solving for the auxiliary field we obtain an Einstein-Hilbert action with modified scalar field interactions. We then proceed to study various specific cases of scalar interactions. In section~\ref{sec:minimal} we analyze cases in which the scalar field is coupled minimally to gravity in the original action. In particular we study the case of a free massive scalar field with just a quadratic potential (section~\ref{sec:minimal:squared}). In contrast to the standard metric formulation of this model we find that in this formulation the resulting potential is that of a well-known inflationary attractor. In section~\ref{sec:nonminimal} we discuss three cases of non-minimally coupled scalars. In particular, in section~\ref{sec:nonminimal:CW} we consider the case of a scalar with scale-invariant interactions. In this model the Einstein-Hilbert term is replaced by the classically scale-invariant coupling $\xi\phi^2R$. Scale-invariance is broken via the Coleman-Weinberg mechanism~\cite{Coleman1973} and the Planck mass is dynamically generated in terms of the VEV of the scalar field. The metric formulation of this model leads to linear inflation~\cite{Kannike2016, Kannike2017a, Artymowski2017, Racioppi2017,  Karam2017, Racioppi2018, Karam2018a}. The Palatini formulation of this model leads to a modified inflationary behaviour with favorable predictions for the corresponding parameters. The slow-roll analysis of the model is carried out in a succeeding section. In section~\ref{sec:nonminimal:induced-gravity} we consider the induced gravity model~\cite{Accetta1985, Kao1990, Carugno1993, Boutaleb-Joutei1997, Freitas1998, Cerioni2009, Tronconi2011, Giudice2014} in which, as in the previous case, the Planck mass is generated by a scalar VEV, determined by a quartic Higgs-like potential. In section~\ref{sec:nonminimal:Higgs} we study the case of nonminimal Higgs inflation model~\cite{DeSimone2009, Barbon2009, Bezrukov2008a, Barvinsky2008,  Barvinsky2009a, Barvinsky2012, Bezrukov2009a, Lerner2010a, Bezrukov2011, Kamada2012, Bezrukov2013, Bezrukov2014, Allison2014, Salvio2015, Hamada2015, Barvinsky2015, Barvinsky2016, Calmet2016, Rubio2018, Enckell2018}, where we find that the $\alpha$-dependence of the Einstein frame potential alleviates the need for large $\xi$, with $\alpha$ being the coefficient of $R^2$. In section~\ref{sec:SR} we consider all the above models in the framework of slow-roll inflation and analyze their predictions. Finally, in the last section we summarize our conclusions.

\section{The Starobinsky model coupled to matter}

Consider the standard Starobinsky action written in terms of an auxiliary scalar $\chi$
\begin{equation}
\mathcal{S}\,=\,\int\!\mathrm{d}^4x\,\sqrt{-g}\left\{\,\frac{1}{2}M_0^2\,R\,+\,\frac{\alpha}{4}R^2\right\}\,=\,\int\,\mathrm{d}^4x\!\sqrt{-g}\left\{\,\frac{1}{2}\left(M_0^2+\,\alpha\chi^2\right)R\,-\frac{\alpha}{4}\chi^4\right\}\ .
\end{equation}
Note that only the bare Planck mass is a dimensionful parameter, $\alpha$ being dimensionless. Aiming at coupling this theory to matter without introducing any other mass scales in the gravity-matter coupling, we introduce a scalar field $\phi$ with a classically scale invariant non-minimal coupling to gravity 
\begin{equation}
\mathcal{S}\,=\,\int\,d^4x\,\sqrt{-g}\left\{\,\frac{1}{2}\left(M_0^2+\,\alpha\chi^2+\xi\phi^2\right)R\,-\frac{1}{2}\left(\nabla\phi\right)^2-\frac{\alpha}{4}\chi^4\,-V(\phi)\,\right\}\ .{\label{JORD}}
\end{equation}
The nonminimal coupling is parametrized by the dimensionless parameter $\xi$. We shall consider this model in the framework of the Palatini formulation, treating the connection $\Gamma$ as an independent variable. Then, the Ricci tensor $R_{\mu\nu}$ is only a function of $\Gamma$. The form of the connection is derived from an additional constraint equation, namely $\delta\mathcal{S}/\delta\Gamma=0$. Therefore, in the Palatini formalism, the connection is an independent variable with no propagating on-shell degrees of freedom, i.e. it is auxiliary.

Next, we consider a Weyl rescaling of the metric
\begin{equation}\label{WEYL}
\bar{g}_{\mu\nu}\,=\,\Omega^2(\phi)\,g_{\mu\nu}\qquad{\text{with}}\qquad\Omega^2(\phi)\,=\,\frac{M_0^2+\xi\phi^2+\alpha\chi^2}{M_P^2}\,.
\end{equation}
The resulting Einstein frame action is
\begin{equation}
 \mathcal{S}\,=\,\int\,\mathrm{d}^4x \sqrt{-\bar{g}}\left\{\,\frac{1}{2}M_P^2\,\bar{R}\,-\frac{1}{2}\frac{\left(\bar{\nabla}\phi\right)^2}{\Omega^2}\,-\bar{V}\right\}\ ,{\label{EINST}}
 \end{equation}
where
\begin{equation}
 \bar{V}(\phi,\chi)\,=\,\frac{1}{\Omega^4}\left(V(\phi)+\frac{\alpha}{4}\chi^4\right)\,.{\label{POT}}
 \end{equation}
As expected, no kinetic term has been generated for the field $\chi$, known as the scalaron in the metric formalism. Note that only the field $\phi$ can act as an inflaton, since it is the only propagating scalar degree of freedom. This is in contrast to the metric formalism where we would obtain two dynamical fields that each can contribute to inflation. The action \eqref{EINST} is standard GR coupled minimally to the scalars $\phi$ and $\chi$, the latter being an auxiliary field. Varying this action with respect to $\Gamma$ produces the usual Levi-Civita expression for the connection, which is written in terms of the metric $\bar{g}_{\mu\nu}$. In what follows, we shall drop the bars on the metric and the gravitational tensors. By varying with respect to $\chi$ we obtain the following constraint equation:
\begin{equation}
\chi^2\,=\,\frac{\frac{4V(\phi)}{(M_0^2+\xi\phi^2)}\,+\,\frac{(\nabla\phi)^2}{M_P^2}}{\left[1-\frac{\alpha(\nabla\phi)^2}{M_P^2(M_0^2+\xi\phi^2)}\right]}\,.{\label{CHI}}
\end{equation}
We may now substitute ({\ref{CHI}}) back into the action. Instead of writing the complete expression of $\mathcal{S}$ in terms of $\chi[\phi]$, we write down an expansion in derivatives neglecting terms $\mathcal{O}((\nabla\phi)^4)$ or higher. We have
\begin{equation}
\mathcal{S}\,\approx\,\int\!d^4x\,\sqrt{-g}\left\{\,\frac{M_P^2}{2}R\,-\frac{1}{2}\frac{(\nabla\phi)^2}{\Omega_0^2}\left(\frac{1}{1+\frac{4\tilde{\alpha} V}{\Omega_0^4}}\right)\,-\frac{V}{\Omega_0^4}\left(\frac{1}{1+\frac{4\tilde{\alpha} V}{\Omega_0^4}}\right)\,+\,\mathcal{O}\left((\nabla\phi)^4\right)\right\}\ ,{\label{ACT-2}}
\end{equation}
where $\tilde{\alpha}=\alpha/M_P^4$ and
\begin{equation}
\Omega_0^2\,=\,\frac{1}{M_P^2}\left(\,M_0^2\,+\,\xi\phi^2\,\right)\,.
\end{equation}
Note that even in the case of minimal coupling $\xi=0$ the presence of the quadratic scalar curvature term modifies the matter Lagrangian non-trivially.

\section{Minimally coupled scalars} 
\label{sec:minimal}

The form of the Palatini-Starobinsky action \eqref{JORD} suggests that even in the case of minimally coupled scalars (i.e. $\xi=0$, $M_0\neq 0$), the presence of the quadratic Starobinsky term is highly non-trivial. Weyl rescaling with $\Omega^2=1+\alpha\chi^2/M_P^2$, using $M_0=M_P$, takes us to the Einstein action \eqref{EINST} with the potential \eqref{POT}. The auxiliary field equation gives
\begin{equation}
\chi^2\,=\,\frac{4V(\phi)+(\nabla\phi)^2}{M_P^2-\alpha\frac{(\nabla\phi)^2}{M_P^2}}
\end{equation}
and the resulting effective action is
\begin{equation}
 \mathcal{S}\,\approx\,\int\,d^4x\,\sqrt{-g}\left\{\frac{M_P^2}{2}R\,-\frac{1}{2}\left(\nabla\phi\right)^2\left(\frac{1}{1+4\tilde{\alpha}V}\right)\,-\frac{V}{\left(1+4\tilde{\alpha}V\right)}\,+\,\mathcal{O}((\nabla\phi)^4)\right\}\ .
 \end{equation}

\subsection{A free massive scalar model}
\label{sec:minimal:squared}

\begin{figure}[H]
\centering
\includegraphics[scale=0.55]{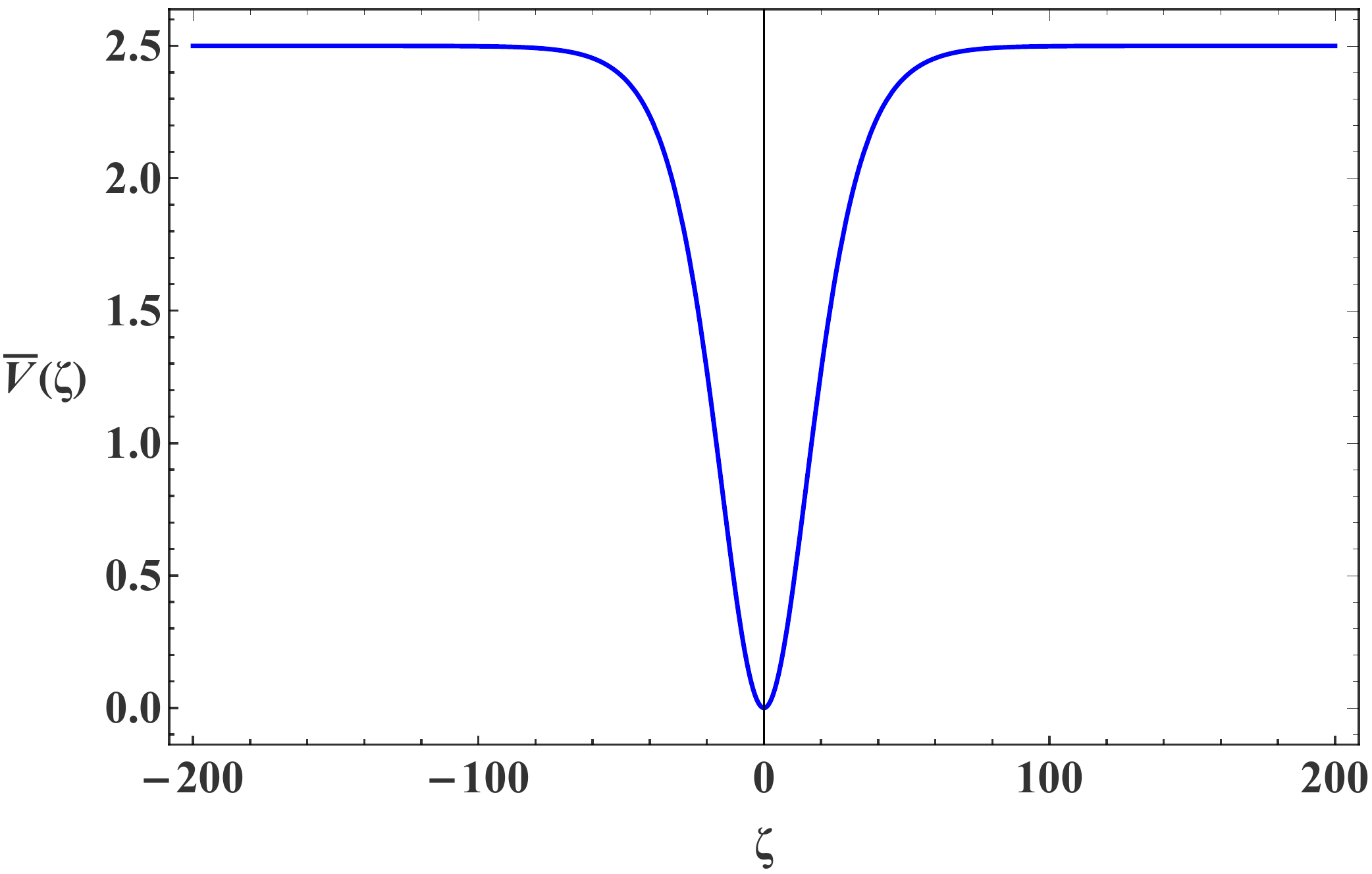}
\caption{The potential for the minimal quadratic model \eqref{eq:pot-squared} as a function of the canonical scalar field given in \eqref{eq:zeta-squared}. We have set $\alpha=0.1$ and $m=0.1$ (in Planck units).}
\label{fig:pot-squared}
\end{figure}

Note that in the simple case of a free massive scalar with a potential
\begin{equation}
 V(\phi)\,=\,\frac{1}{2}m^2\phi^2\ ,
 \end{equation}
we can obtain an inflationary plateau in the effective theory described above. Indeed, by introducing the canonical scalar
\begin{equation}
\zeta\,=\,\int\frac{d\phi}{\sqrt{1+\frac{2\alpha m^2}{M_P^4}\phi^2}}\,=\,\frac{M_P^2}{m\sqrt{2\alpha}}\sinh^{-1}(m\sqrt{2\alpha}\phi/M_P^2)\ ,
\label{eq:zeta-squared}
\end{equation}
or
\begin{equation}
 \phi\,=\,\frac{M_P^2}{m\sqrt{2\alpha}}\sinh\left(m\zeta\sqrt{2\alpha}/M_P^2\right)\,,
 \end{equation}
the resulting potential is
\begin{equation}
\bar{V}(\zeta)\,=\,\frac{M_P^2}{4\alpha}\,\tanh^2\left({m\sqrt{2\alpha}\zeta}/{M_P^2}\right)\ .
\label{eq:pot-squared}
\end{equation}
This is a well-known potential, corresponding to an inflationary attractor, also obtainable in supergravity in the framework of the $SU(2,1)/SU(2)\times U(1)$ no-scale model with a K\"{a}hler potential $-3\ln(T+\bar{T}-|S|^2)$ and a superpotential $W=W_0+S(T-1)/(T+1)$~\cite{Kallosh2013b, Lahanas2015, Carrasco2015a, Carrasco2015}. In section~\ref{sec:SR:squared} we analyze numerically the predicted slow-roll parameters and find them in full agreement with observations for a wide range of parameter choices. Large values of $r$ can be avoided by reasonably small values of the parameter $\alpha$. This is in sharp contrast to the case of the quadratic model in chaotic inflation. A plot of the potential \eqref{eq:pot-squared} is shown in Fig.~\ref{fig:pot-squared}.

We note that even in the case of a minimally coupled scalar with a quartic Higgs-like potential $\frac{\lambda}{4}(\phi^2-v^2)^2$, the presence of the $R^2$ term has a non-trivial effect on the Einstein-frame potential. In this case, for large values of the canonical field the effective potential $V/(1+4\tilde{\alpha}V)$ reaches an inflationary plateau $M_P^4/4\alpha$.

\section{Non-Minimally coupled scalars}
\label{sec:nonminimal}

\subsection{A quasi-scale-invariant framework}
\label{sec:nonminimal:CW}
 
The fact that the $R^2$ term of the Starobinsky model dominates over the linear term during inflation suggests that at very high energies gravity may be scale invariant. Adopting an effective theory approach and hoping that the results are general enough for some possible UV completions, we proceed considering a quasi-scale-invariant version of action \eqref{JORD}, meaning that $M_0=0$ and no other dimensionful parameters are present apart from a cosmological constant term\footnote{We allow for the possibility of a bare cosmological constant term needed to ensure vanishing potential at the minimum.} $\Lambda^4$, the scalar potential being the quartic $V(\phi)=\frac{\lambda}{4}\phi^4$
\begin{equation}
\mathcal{S}\,=\,\int\,d^4x\,\sqrt{-g}\,\left\{\,\frac{1}{2}\left(\xi\phi^2+\alpha\chi^2\right)R\,-\frac{1}{2}\left(\nabla\phi\right)^2\,-\frac{\lambda}{4}\phi^4\,-\frac{\alpha}{4}\chi^4\,+\,\Lambda^4\,\right\}{\label{JORD-1}}\ .
\end{equation}
To this action we should add the matter action $ \mathcal{S}_m\,=\,\int\,d^4x\,\sqrt{-g}\,\mathcal{L}_m(\varphi',\,\psi,\,A_{\mu})$, 
containing all matter fields interacting with $\phi$ through scale-invariant interactions. These interactions at the quantum level will generate at one-loop level radiative corrections, which, calculated in the Jordan frame, are grouped in an effective potential $V_1(\phi)$. Assuming that the detailed field content of $\mathcal{S}_m$ is such that the Coleman-Weinberg mechanism~\cite{Coleman1973} takes place, the one-loop effective potential has the form~\cite{Kannike2016, Kannike2017a, Artymowski2017, Racioppi2017, Racioppi2018, Karam2017, Karam2018a}\footnote{See~\cite{Shaposhnikov2009a, Garcia-Bellido2011, Khoze2013, Bezrukov2013a, Gabrielli2014, Salvio2014, Csaki2014a, Kannike2015, Marzola2016a, Barrie2016, Ferreira2016, Marzola2016b, Rinaldi2016a, Farzinnia2016, Karananas2016, Tambalo2017, Kannike2017, Ferreira2017, Salvio2017, Kannike2017a, Barnaveli2018, Ferreira2018} for other inflationary models based on classical scale invariance.}
\begin{equation}
V_1(\phi)\,=\,\Lambda^4\left(\,1\,+\,\frac{\phi^4}{\langle\phi\rangle^4}\left(\,2\ln(\phi^2/\langle\phi\rangle^2)\,-1\right)\right)\,=\,\Lambda^4\left(\,1\,+\,\frac{\xi^2\phi^4}{M_P^4}\left(\,2\ln(\xi\phi^2/M_P^2)\,-1\right)\,\,\right)\,,{\label{ONE-LOOP}}
\end{equation}
where the Planck mass is defined in terms of the VEV of $\phi$ that minimizes $V_1$, namely $M_P^2\equiv\,\xi\langle\phi\rangle^2$. Thus, after dimensional transmutation, the quartic $\phi^4$ term is replaced by ({\ref{ONE-LOOP}}).

The cosmological constant $\Lambda$ is required in order to ensure the vanishing of the potential $V_1(\phi)$ at the minimum. Alternatively, we may replace this condition with the vanishing of the overall resulting potential $\bar{V}(\phi,\chi)\propto V_1(\phi)+\alpha\chi^4/4$. In this case $V_1(\phi)$ would be just
$V_1(\phi)\,=\,C\,\phi^4\left(2\ln(\phi^2/\langle\phi\rangle^2)-1\right)$,
where $C$ is a coefficient dependent on the details of the matter action $\mathcal{S}_m$. Then, the vanishing of $\bar{V}$ would have to be attributed to a tuned cancellation between the two terms.

Considering now the Weyl rescaling ({\ref{WEYL}}) we obtain again the Einstein frame action
\begin{equation} 
\mathcal{S}\,=\,\int\,\mathrm{d}^4x\sqrt{-\bar{g}}\left\{\,\frac{1}{2}M_P^2\,\bar{R}\,-\frac{1}{2}\frac{\left(\bar{\nabla}\phi\right)^2}{\Omega^2}\,-\bar{V}\right\}\ ,{\label{EINST-1}}
\end{equation}
with
\begin{equation}
\bar{V}(\phi,\chi)\,=\,\frac{1}{\Omega^4}\left(V_1(\phi)+\frac{\alpha}{4}\chi^4\right)\,
\end{equation}
and $\Omega^2=\left(\xi\phi^2+\alpha\chi^2\right)/M_P^2$.
The equation for the auxiliary field $\chi$ is
\begin{equation}
\chi^2\,=\,\frac{4V_1(\phi)\,+\,\frac{\xi\phi^2(\nabla\phi)^2}{M_P^2}}{\left[\xi\phi^2-\frac{\alpha(\nabla\phi)^2}{M_P^2}\right]}\,.{\label{CHI-1}}
\end{equation}
and, substituted in ({\ref{EINST-1}}) gives up to $\mathcal{O}((\nabla\phi)^4)$ the effective action
\begin{equation}
\mathcal{S}\,\approx\,\int\,d^4x\,\sqrt{-g}\left\{\,\frac{M_P^2}{2}R\,-\frac{1}{2}(\nabla\phi)^2\left(\frac{\xi\phi^2M_P^2}{\xi^2\phi^4+4{\alpha} V_1}\right)\,-V_1\left(\frac{M_P^4}{\xi^2\phi^4+4{\alpha} V_1}\right)\,+\,\mathcal{O}((\nabla\phi)^4)\right\}\ .{\label{ACT-3}}
\end{equation}
Note that in the absence of radiative corrections, i.e. just for a $\lambda\phi^4/4$ potential, we would trivially get a constant potential, while the kinetic term of $\phi$ would be rescaled by a trivial $\alpha$-dependent factor. 
In the absence of the Starobinsky term, i.e. for $\alpha=0$, it has been shown that the Coleman-Weinberg action \eqref{ACT-3} leads to linear inflation~\cite{Racioppi2017} for $\xi \gtrsim 0.1$. 

\begin{figure}[H]
\centering
\includegraphics[scale=0.45]{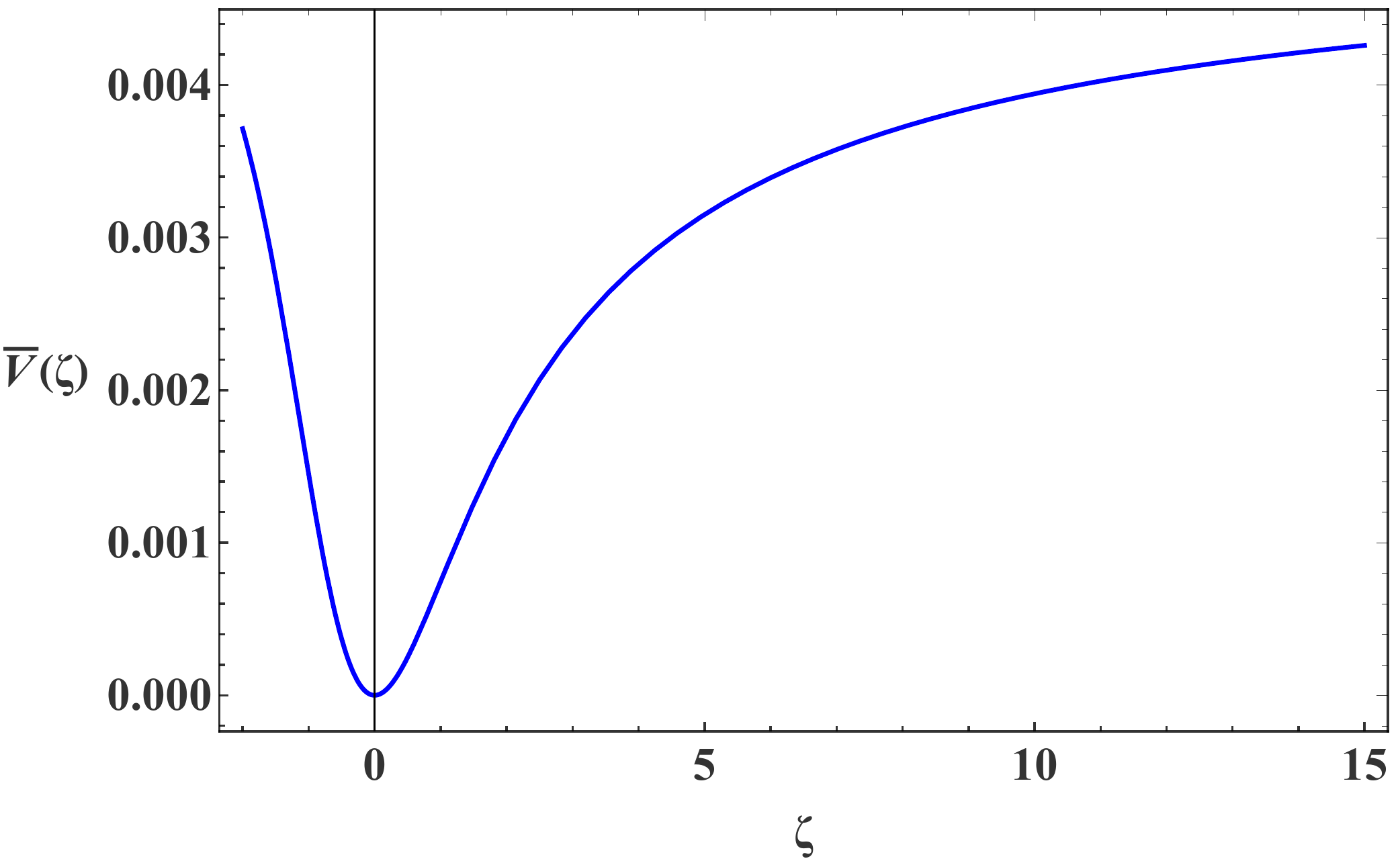}
\caption{The potential for the nonminimal Coleman-Weinberg potential \eqref{eq:pot-CW} as a function of the canonical scalar field given in \eqref{eq:zeta-CW}. We have set $\Lambda=10^{-2}$, $\xi=0.1$ and $\alpha=0.01$ (in Planck units).}
\label{fig:pot-CW}
\end{figure}

For a general $\alpha$ we may introduce the canonical scalar field variable
\begin{equation}
\zeta\,=\,M_P\int\,d\phi\,\sqrt{\frac{\xi\phi^2}{\xi^2\phi^4+4\alpha V_1(\phi)}} \ .
\end{equation}
The parameter $\Lambda$ giving the overall scale of the one loop potential is estimated to be a few orders of magnitude below $M_P$~\cite{Kannike2016}. Therefore, we may approximate the canonical scalar to be 
\begin{equation}
\zeta\,\approx\,\frac{M_{P}}{2\sqrt{\xi}}\ln(\xi\phi^2/M_P^2)\,.
\label{eq:zeta-CW}
\end{equation}
The effective Lagrangian is 
\begin{equation}
 \mathcal{L}_{\text{eff}}\,\approx\,-\frac{1}{2}\left(\nabla\zeta\right)^2\,- \bar{V}(\zeta)\ ,{\label{CSI-CW}}
\end{equation}
where
\begin{equation}
\bar{V}(\zeta) = \frac{\Lambda^4\left(4\sqrt{\xi}\frac{\zeta}{M_P}\,-1\,+\,e^{-4\sqrt{\xi}\frac{\zeta}{M_P}}\right)}{M_P^4+4\alpha\Lambda^4\left(4\sqrt{\xi}\frac{\zeta}{M_P}\,-1\,+\,e^{-4\sqrt{\xi}\frac{\zeta}{M_P}}\right)} \ .
\label{eq:pot-CW}
\end{equation}
For large values of $\zeta$ values there is an inflationary plateau of height $1/4\alpha$, while the potential minimum occurs at $\zeta=0$. A plot of the potential \eqref{eq:pot-CW} is given in Fig.~\ref{fig:pot-CW}. 

\subsection{Induced gravity model}
\label{sec:nonminimal:induced-gravity}

In this section we study another interesting model that can dynamically generate the Planck mass from the VEV of a scalar field. The dynamics of the scalar field are described by a Higgs-like potential with a minimum at the scalar VEV, $v\equiv\braket{\phi}$. Such models~\cite{Accetta1985, Kao1990, Carugno1993, Boutaleb-Joutei1997, Freitas1998, Cerioni2009, Tronconi2011, Giudice2014} have their origin in early attempts to marry the dynamics of spontaneous symmetry breaking and gravity, but nowadays they serve as viable extensions of GR, at least when one studies inflation.

\begin{figure}[H]
\centering
\includegraphics[scale=0.5]{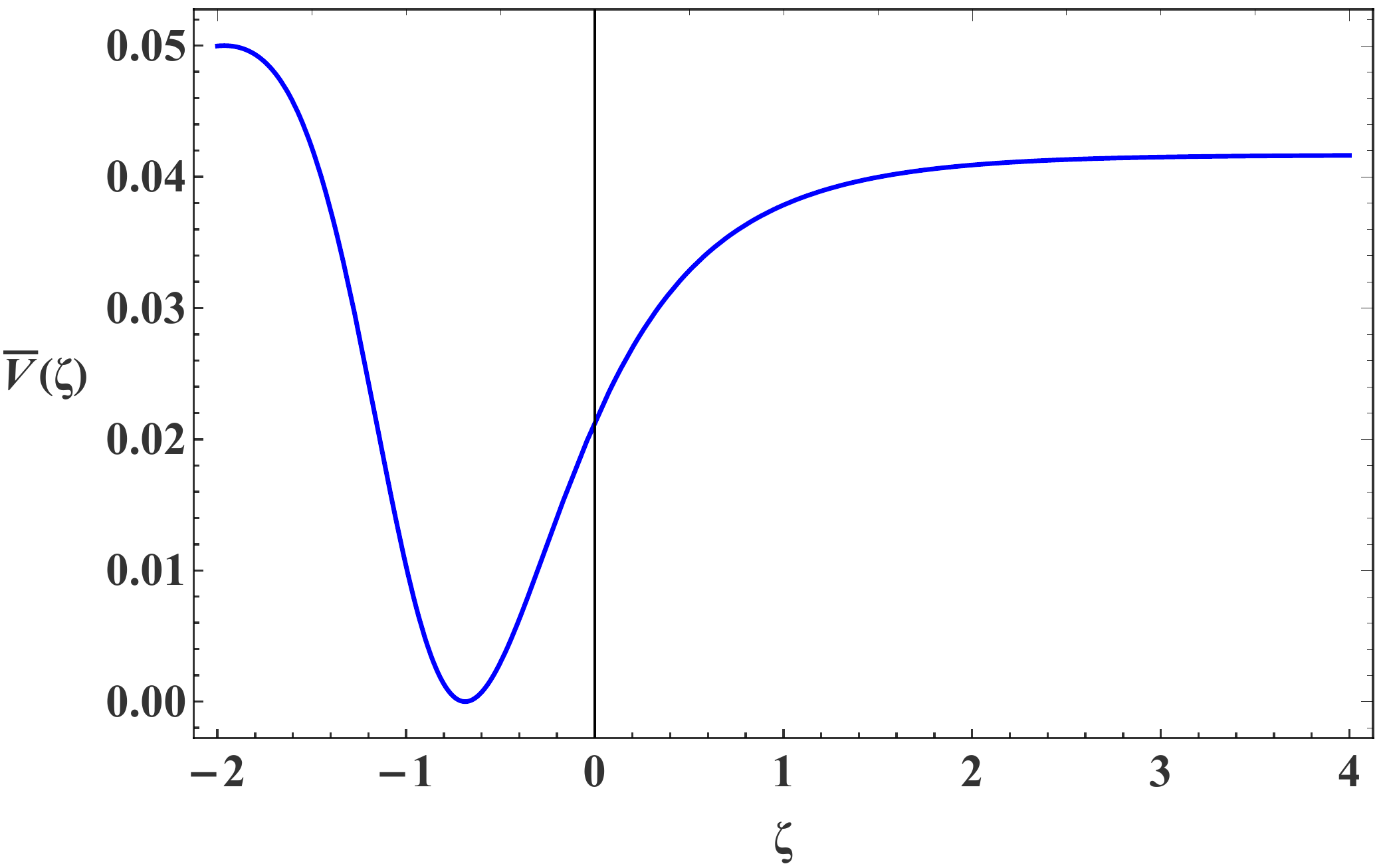}
\caption{The potential for the induced gravity model \eqref{eq:pot-induced} as a function of the canonical scalar field given in \eqref{eq:zeta-induced}. We have set $\lambda=10^{-2}$, $\xi=0.1$, $\alpha=0.01$ and $\upsilon=1/\sqrt{\xi}$ (in Planck units).}
\label{fig:pot-induced}
\end{figure}

Consider the following action
\begin{equation}
\mathcal{S}=\int\!\mathrm{d}^4x\,\sqrt{-g}\left\{\frac{1}{2}\left(\xi\phi^2+\alpha\chi^2\right)R-\frac{1}{2}(\nabla\phi)^2-\frac{\lambda}{4}\left(\phi^2-v^2\right)^2-\frac{\alpha}{4}\,\chi^4\right\}\,.
\end{equation}
Following what we did previously, using \eqref{CHI-1}, we end up with the effective action in the Einstein frame
\begin{equation}
\mathcal{S}\approx\int\!\mathrm{d}^4x\,\sqrt{-g}\left\{\frac{M_P^2}{2}\,R-\frac{1}{2}(\nabla\phi)^2\left(\frac{\xi\phi^2M_P^2}{\xi^2\phi^4+4\alpha V}\right)-V\left(\frac{M_P^4}{\xi^2\phi^4+4\alpha V}\right)+\mathcal{O}((\nabla\phi)^4)\right\},
\end{equation}
where $V(\phi)=\frac{\lambda}{4}(\phi^2-v^2)^2$ is the induced gravity potential. We may introduce the canonical scalar field
\begin{equation}
\zeta=\frac{M_P}{2}\sqrt{\frac{\xi}{\xi^2+\alpha\lambda}}\ln{\left[\phi^2(\alpha\lambda+\xi^2)-\alpha\lambda v^2+\sqrt{\alpha\lambda+\xi^2}\sqrt{\xi^2\phi^4+\alpha\lambda(\phi^2-v^2)^2}\right]}\ ,
\label{eq:zeta-induced}
\end{equation}
and the potential, expressed in terms of the canonical field reads as
\begin{equation}
\bar{V}=M_P^4\,\frac{V}{\xi^2\phi^4+4\alpha V}=\frac{\lambda M_P^4}{4(\xi^2+\alpha\lambda)}\left(\frac{e^{\frac{4\sqrt{\alpha\lambda+\xi^2}\zeta}{M_P\sqrt{\xi}}}-2e^{\frac{2\sqrt{\alpha\lambda+\xi^2}\zeta}{M_P\sqrt{\xi}}}-\alpha\lambda M^4_P}{\alpha\lambda M^4_P + e^{\frac{4\sqrt{\alpha\lambda+\xi^2}\zeta}{M_P\sqrt{\xi}}}}\right)^2 \ ,
\label{eq:pot-induced}
\end{equation}
where we used $v^2=M^2_P/\xi$. A plot of the potential \eqref{eq:pot-induced} is given in Fig.~\ref{fig:pot-induced}.

\subsection{Nonminimal Higgs model}
\label{sec:nonminimal:Higgs}

A compelling idea concerning inflation, is for the Higgs boson to take up the role of the inflaton field~\cite{DeSimone2009, Barbon2009, Bezrukov2008a, Barvinsky2008,  Barvinsky2009a, Barvinsky2012, Bezrukov2009a, Lerner2010a, Bezrukov2011, Kamada2012, Bezrukov2013, Bezrukov2014, Allison2014, Salvio2015, Hamada2015, Barvinsky2015, Barvinsky2016, Calmet2016, Rubio2018, Enckell2018}. This can by achieved by coupling the Higgs field nonminimally to the Ricci curvature. Such a proposal is minimal in the sense that one does not need to extend the field content of the Standard Model. There have been numerous studies through the years, with the general consensus being that one has to expect very large values of the nonminimal coupling $\xi\sim\left(10^3-10^5\right)$, in order to account for the measured values of the inflationary observables. Here and in following sections we argue that in the Palatini formalism the nonminimal coupling $\xi$ can take up much lower values.

\begin{figure}[H]
\centering
\includegraphics[scale=0.55]{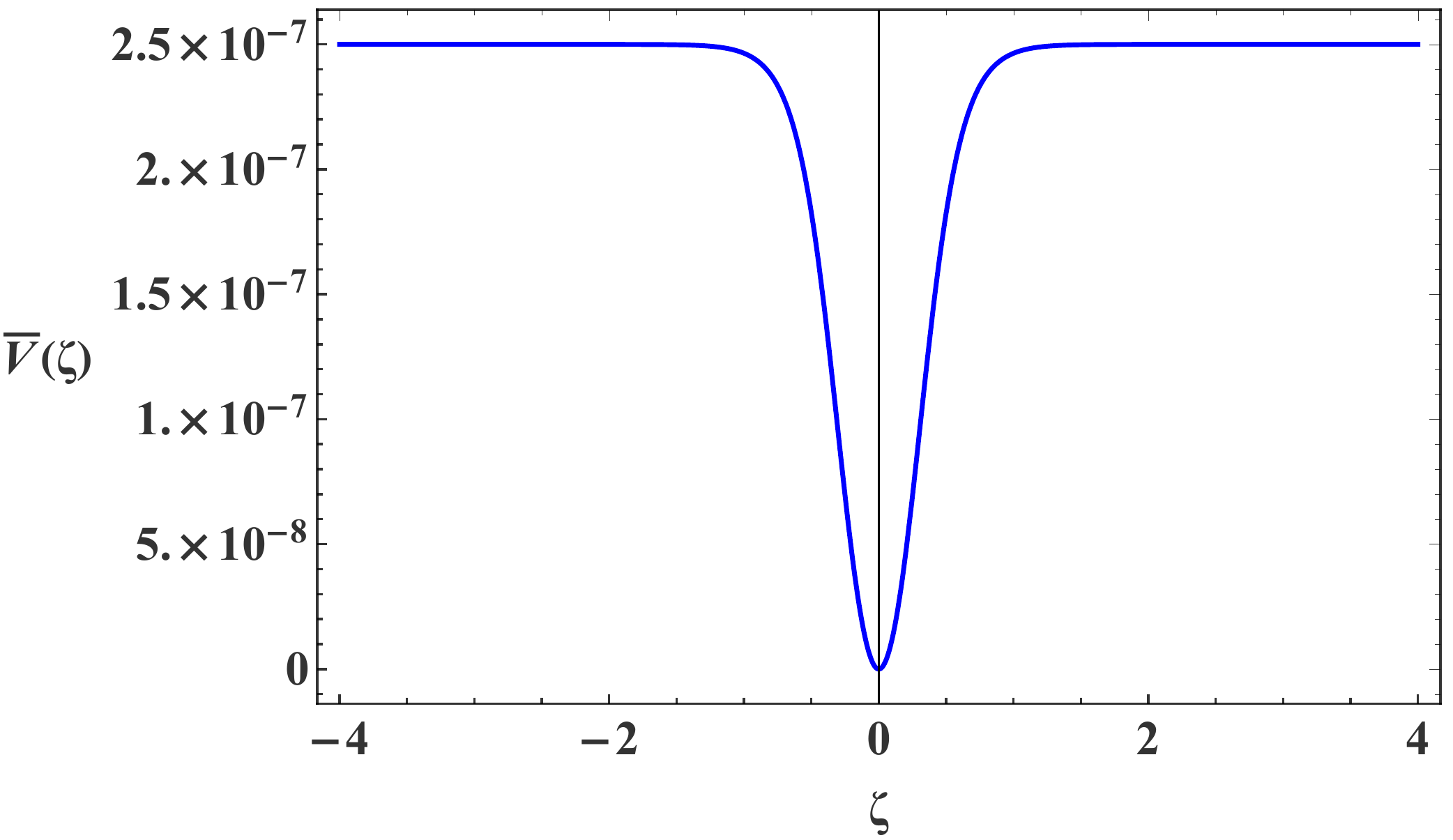}
\caption{The potential for the nonminimal Higgs potential \eqref{eq:pot-higgs} as a function of the canonical scalar field given in \eqref{eq:zeta-higgs}. We have set $\lambda=10^{-4}$, $\xi=10$ and $\alpha=1$.}
\label{fig:pot-higgs}
\end{figure}

Consider the following action
\begin{equation}
\mathcal{S}=\int\!\mathrm{d}^4x\sqrt{-g}\left\{\frac{1}{2}\left(M_0^2+2\xi\left|H\right|^2+\alpha\chi^2\right)R-\frac{1}{2}\left|DH\right|^2-V(H)-\frac{\alpha}{4}\chi^4\right\}\,.
\end{equation}
By performing the Weyl rescaling of the metric with
\begin{equation}
\Omega^2=1+\frac{\alpha\chi^2}{M_P^2}+2\xi\,\frac{\left|H\right|^2}{M_P^2}
\end{equation}
the Einstein frame action reads
\begin{equation}
\mathcal{S}=\int\!\mathrm{d}^4x\sqrt{-g}\left\{\frac{M_P^2}{2}R-\left|DH\right|^2\frac{1}{\left(1+\frac{\alpha\chi^2}{M_P^2}+2\xi\,\frac{\left|H\right|^2}{M_P^2}\right)}-\frac{V(H)+\frac{\alpha}{4}\,\chi^4}{\left(1+\frac{\alpha\chi^2}{M_P^2}+2\xi\,\frac{\left|H\right|^2}{M_P^2}\right)^2}\right\}\,,
\end{equation}
where we assumed that $M_0\simeq M_P$. Adopting the unitary gauge $H=\frac{1}{\sqrt{2}}\begin{pmatrix}
0\\h
\end{pmatrix}$ with
\begin{equation}
V(H)=\lambda\left(\left|H\right|^2-\frac{v^2}{2}\right)^2=\frac{\lambda}{4}\left(h^2-v^2\right)^2
\end{equation}
and
\begin{equation}
\chi^2=\frac{\frac{4V}{M_0^2+2\xi\left|H\right|^2}+\frac{2\left|\nabla H\right|^2}{M_P^2}}{\left[1-\frac{2\alpha\left|\nabla H\right|^2}{M_P^2\left(M_0^2+2\xi\left|\nabla H\right|^2\right)}\right]}\,,
\end{equation}
we obtain
\begin{equation}
\mathcal{S}=\int\!\mathrm{d}^4x\sqrt{-g}\left\{\frac{M_P^2}{2}\,R-\frac{1}{2}(\nabla h)^2\left(\frac{\xi h^2 M_P^2}{\xi^2 h^4+4\alpha V}\right)-V\left(\frac{M_P^4}{\xi^2 h^4+4\alpha V}\right)+\mathcal{O}\left((\nabla h)^4\right)\right\}\ .
\end{equation}
Next, we assume that in order for the Higgs field to drive inflation it must be far away from its VEV. We also assume that $\xi h^2\gg M_P^2$ and obtain the following expression for the canonically normalized field
\begin{equation}
\zeta \simeq M_P \sqrt{\frac{\xi}{\xi^2 + \alpha\lambda}} \sinh^{-1} \left( \frac{h}{M_P} \sqrt{\xi\, \frac{\xi^2 + \alpha\lambda}{\xi^2 - \alpha\lambda}} \right)\ .
\label{eq:zeta-higgs}
\end{equation}
The potential reads as
\begin{equation}
\bar{V}(\zeta) \simeq \frac{\lambda}{4} \frac{M^4_P}{\xi^2 + \alpha\lambda} \frac{\sinh^2 \left( \sqrt{\frac{\xi^2 + \alpha\lambda}{\xi} } \frac{\zeta}{M_P} \right)}{\frac{2\xi^2}{\xi^2 - \alpha\lambda} + \sinh^2\left( \sqrt{\frac{\xi^2 + \alpha\lambda}{\xi} } \frac{\zeta}{M_P} \right)}\ .
\label{eq:pot-higgs}
\end{equation}
Note that the nonminimal coupling $\xi$ has been replaced with an effective combination that depends on $\alpha$. As the slow-roll inflation analysis will show below, this can potentially extend the viability of the model for reasonably small values of $\xi$.
A plot of the above potential is shown in Fig.~\ref{fig:pot-higgs}. 

\section{Slow-Roll Inflation} 
\label{sec:SR}

The Einstein equation resulting from the general form of the action ({\ref{ACT-2}}) is
\begin{equation}
 M_P^2\left(R_{\mu\nu}\,-\frac{1}{2}g_{\mu\nu}R\right)\,=\,\frac{\nabla_{\mu}\phi\nabla_{\nu}\phi}{\Omega_0^2}\left(\frac{1}{1+\frac{4\tilde{\alpha} V}{\Omega_0^4}}\right)\,-g_{\mu\nu}\left[\frac{1}{2}\frac{(\nabla\phi)^2}{\Omega_0^2}\left(\frac{1}{1+\frac{4\tilde{\alpha} V}{\Omega_0^4}}\right)\,+\,\frac{V}{\Omega_0^4}\left(\frac{1}{1+\frac{4\tilde{\alpha} V}{\Omega_0^4}}\right)\,\,\right]\ .
 \end{equation}
Assuming a flat FLRW metric $ds^2\,=\,-dt^2\,+\,a^2(t)(d\vec{x})^2$, the resulting Friedmann equation reads
\begin{equation}
3M_P^2H^2\,=\,\left[\frac{1}{2}\frac{(\dot{\phi})^2}{\Omega_0^2}\,+\,\frac{V}{\Omega_0^4}\,\right]\left(\frac{1}{1+\frac{4\tilde{\alpha} V}{\Omega_0^4}}\right)\ ,
\end{equation}
with $H=\dot{a}/a$ the usual Hubble parameter. Equivalently, the action can be expressed in terms of the canonically normalized field $\zeta$ and the potential $\bar{V}(\zeta)$ takes the form
\begin{equation}
 \mathcal{S}\,\approx\,\int\,d^4x\,\sqrt{-g}\left\{\,-\frac{1}{2}(\nabla\zeta)^2-\bar{V}(\zeta)\right\}\ ,
 \end{equation}
where
\begin{equation}
\zeta=\int\frac{d\phi}{\sqrt{\Omega_0^2+\frac{4\tilde{\alpha}V(\phi)}{\Omega_0^2}}}\,,\,\,\,\,\,\bar{V}(\zeta)\,=\,\frac{V(\phi)}{\Omega_0^4+4\tilde{\alpha}V(\phi)}\ .
\end{equation}
Then, the Friedmann equation has its canonical form
\begin{equation}
3M_P^2H^2\,=\,\frac{1}{2}\dot{\zeta}^2\,+\,\bar{V}(\zeta)\ .
\end{equation}

In the slow-roll approximation, the first order expressions for the tensor-to-scalar ratio $r$ and the scalar index $n_s$ are given in terms of the potential slow-roll parameters $\epsilon_V$ and $\eta_V$ at horizon crossing as 
\begin{equation}
 r\,\approx\,16\,\epsilon_V\ ,\qquad n_s\,\approx\,1-6\epsilon_V\,+\,2\eta_V\ ,
 \end{equation}
where
\begin{equation}
 \epsilon_V\,\equiv\,\frac{M_P^2}{2}\left(\frac{\bar{V'}(\zeta)}{\bar{V}(\zeta)}\right)^2\,, \qquad \eta_V\,\equiv\,M_P^2\left(\frac{\bar{V''}(\zeta)}{\bar{V}(\zeta)}\right)\ .
 \end{equation}
Next, we compute these parameters for each of the models analyzed in the preceding sections.

\subsection{The minimally-coupled quadratic model}
\label{sec:SR:squared}

The slow-roll parameters for the minimally-coupled quadratic model, described by the potential \eqref{eq:pot-squared}, are found to be
\begin{equation}
 \epsilon_V\,=\,\frac{m^2}{ M_P^2}\frac{16\alpha}{\sinh^2(2m\sqrt{2\alpha}\zeta/M_P^2)}\ ,
 \end{equation}
\begin{equation}
 \eta_V\,=\,16\alpha\frac{m^2}{ M_P^2}\frac{\left(2-\cosh(2m\zeta\sqrt{2\alpha}/ M_P^2)\right)}{\sinh^2(2m\zeta\sqrt{2\alpha}/M_P^2)}\ .
\end{equation}
The end of inflation condition $\epsilon_V\simeq 1$ gives
\begin{equation}
\zeta_f\,=\,\frac{M_P^2}{2m\sqrt{2\alpha}}\sinh^{-1}(2m\sqrt{2\alpha}/M_P)\ .
\end{equation}
Then, the integral for the number of e-folds in terms of the canonical scalar field yields
\begin{equation}
N\,=\,\int_{\zeta_*}^{\zeta_f}\frac{d\zeta}{\sqrt{2\epsilon_V(\zeta)}}\,=\,\frac{M^2_P}{16\alpha m^2}\left.\int\frac{dx}{x}\sqrt{\sinh x}\right|_{2m\zeta_*\sqrt{2\alpha}/M_P^2}^{2m\zeta_f\sqrt{2\alpha/}M_P^2}\ .
\end{equation}

\begin{figure}[H]
\centering
\includegraphics[width=.55\textwidth]{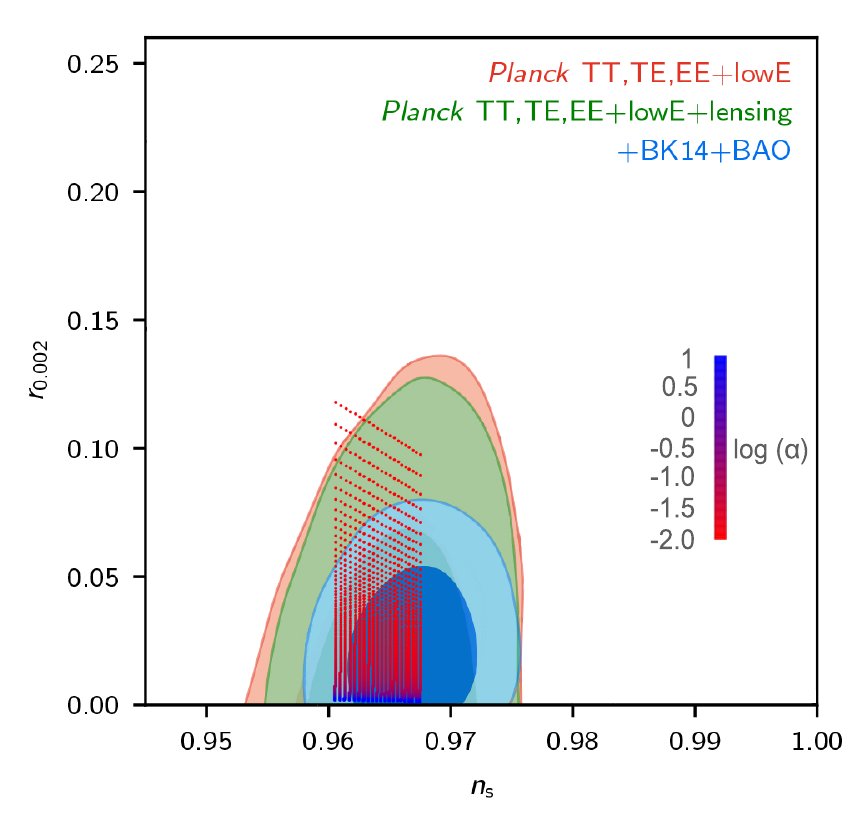}
\caption{The predictions for the inflationary observables in the $n_S-r$ plane for the minimally-coupled quadratic model for $N=50-60$ e-folds. We have set $m=0.1$ and have varied $\alpha$ between $0.01$ and $10$.}
\label{fig:r_ns-squared}
\end{figure}

In Fig.~\ref{fig:r_ns-squared} we plot the predictions of the model in the $n_s-r$ plane for various values of the parameter $\alpha$ and for $N=50-60$ e-folds, overlaid with the latest results from the Planck 2018 collaboration~\cite{Aghanim2018, Akrami2018}. We observe that as $\alpha$ becomes smaller, the predictions asymptote to those of the simple quadratic model (without the $R^2$ term). On the contrary, as $\alpha$ becomes larger the value of $r$ becomes smaller, reaching values similar to those of the Starobinsky model. 

\subsection{The nonminimal Coleman-Weinberg model}
\label{sec:SR:CW}

A direct calculation of $\epsilon_V$ and $\eta_V$ for the nonminimal Coleman-Weinberg potential in \eqref{eq:pot-CW} gives
\begin{equation}
 \epsilon_V\,\approx\,\frac{M_P^2}{2\left(\zeta-M_P/4\sqrt{\xi}\right)^2\left(1+16\alpha\sqrt{\xi}\frac{\Lambda^4}{M_P^5}\left(\zeta-M_P/4\sqrt{\xi}\right)\right)^2}   
\end{equation}
and
\begin{equation}
 \eta_V\,\approx\,-\frac{32\alpha\sqrt{\xi}\frac{\Lambda^4}{M_P^4}}{\left(\zeta-M_P/4\sqrt{\xi}\right)^2\left(1+16\alpha\sqrt{\xi}\frac{\Lambda^4}{M_P^4}\left(\zeta-M_P/4\sqrt{\xi}\right)\right)^2}\ .
 \end{equation}
From the expression for $\epsilon_V$ we may determine the final value $\zeta_f$ corresponding to the end of inflation as defined by the condition $\epsilon_V\simeq1$. It is
\begin{equation}
 \zeta_f\,=\,4\sqrt{\xi}M_P\,+\,\frac{M_P^4}{32\alpha\sqrt{\xi}\Lambda^4}\left(-1+\sqrt{1+\frac{\sqrt{2}M_P^4}{16\alpha\sqrt{\xi}\Lambda^4}}\right)\,\approx\,4\sqrt{\xi}M_P+M_P/2\sqrt{2}-2\alpha\sqrt{\xi}\frac{\Lambda^4}{M_P^3}\ .
 \end{equation}
Inserting the approximate expression for $\epsilon_V$ in the definition of the number of e-folds we obtain
\begin{equation}
 N\,=\,\ln\left(\zeta_*/M_P-4\sqrt{\xi}\right)\,-\ln\left(\zeta_f/M_P-4\sqrt{\xi}\right)\,+\,16\alpha\sqrt{\xi}\frac{\Lambda^4}{M_P^5}\left(\zeta_*-\zeta_f\right)\ .
 \end{equation}

\begin{figure}[h]
\centering
\includegraphics[width=0.6\textwidth]{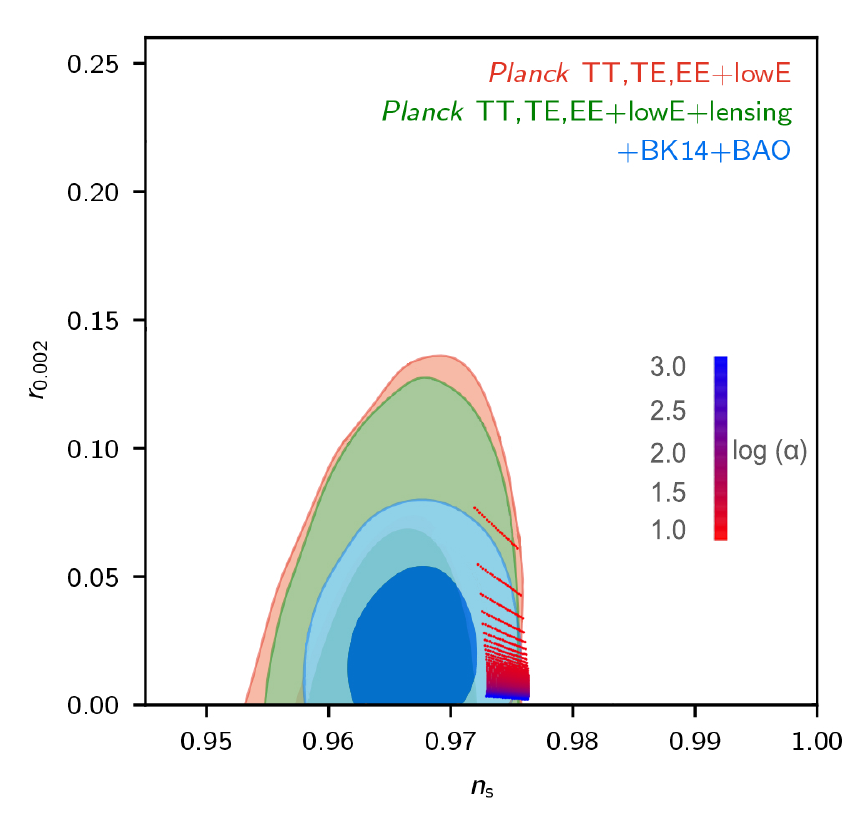}
\caption{The predictions for the inflationary observables in the $n_S-r$ plane for the nonminimal Coleman-Weinberg model and for $N=50-60$ e-folds. We have set $\Lambda = 0.1$ and $\xi = 0.1$ (in Planck units).}
\label{fig:r_ns-CW}
\end{figure}

In Fig.~\ref{fig:r_ns-CW} we plot the predictions of the model in the $n_s-r$ plane for various values of the parameter $\alpha$, with fixed $\Lambda = 0.1$, $\xi = 0.1$ and for $N=50-60$ e-folds. We observe that as $\alpha$ becomes smaller, the predictions asymptote to those of the linear inflation model~\cite{Kannike2016, Artymowski2017, Racioppi2017, Karam2017, Kannike2017a, Racioppi2018, Karam2018a} (without the $R^2$ term)\footnote{In the model without the $R^2$ term, the linear inflation limit is reached for values of the nonminimal coupling above $\xi \gtrsim 0.1$}. As $\alpha$ becomes larger the value of $r$ becomes smaller since the $R^2$ term dominates.

\subsection{The induced gravity model}
\label{sec:SR:induced} 

\begin{figure}[H]
\includegraphics[width=.5\textwidth]{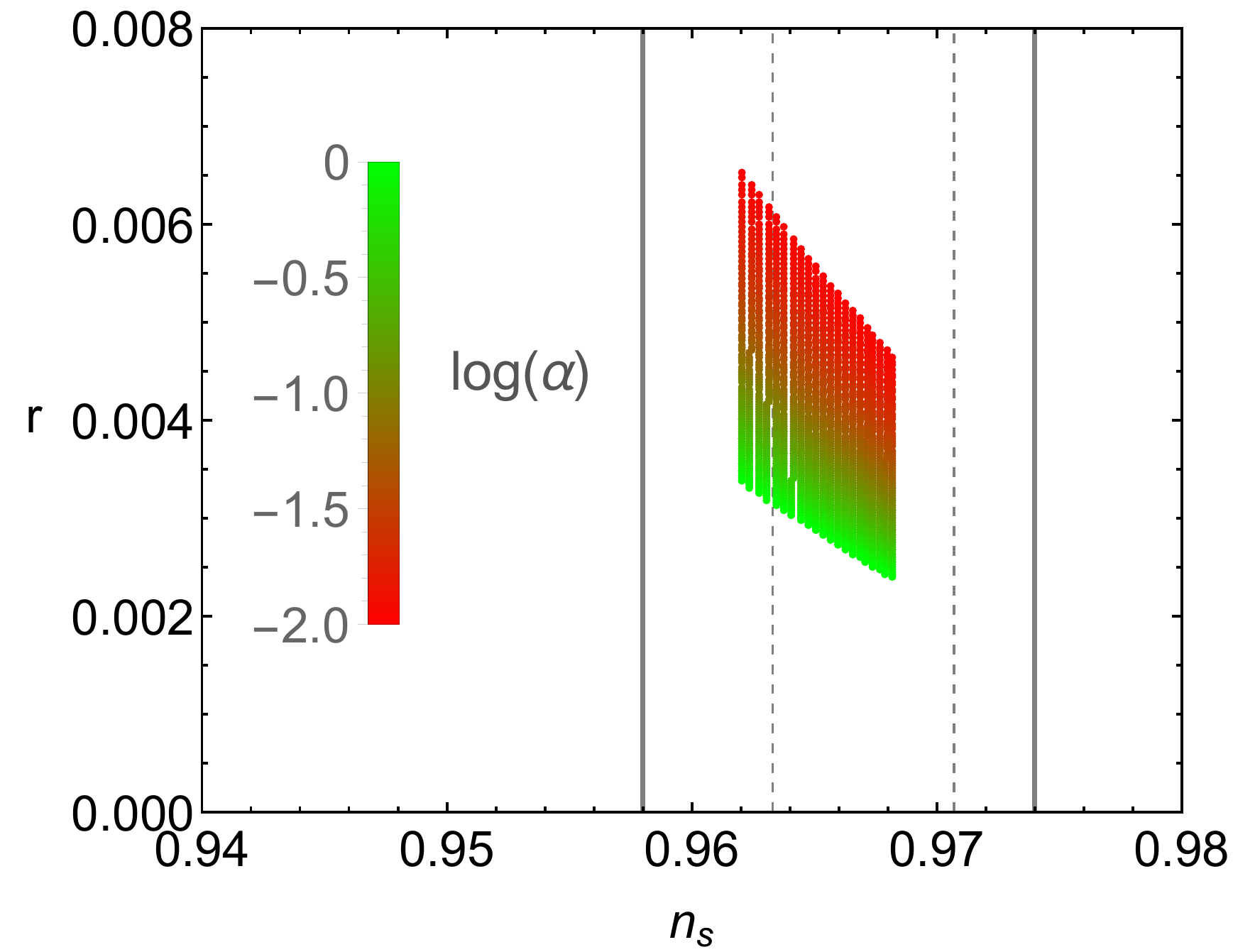}
\includegraphics[width=.5\textwidth]{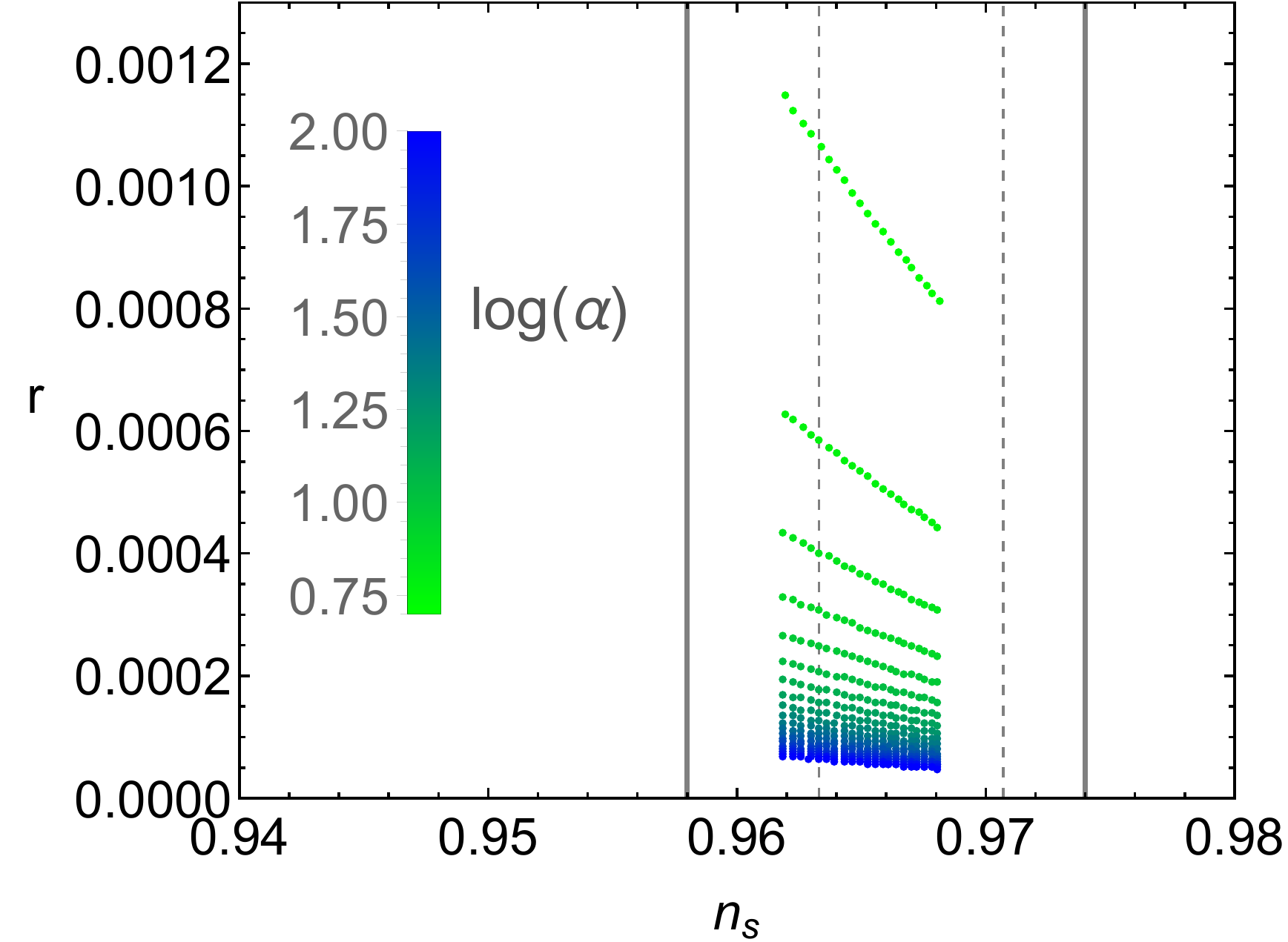}
\caption{The predictions for the inflationary observables in the $n_S-r$ plane for the induced gravity model and for $N=50-60$ e-folds. In the plot on the left we have varied $\alpha$ between $0.1$ and $10$, while in the plot on the right we have varied $\alpha$ between $10$ and $100$. We have set $\lambda = 0.01$, $\xi = 0.1$ (in Planck units).}
\label{fig:r_ns-induced}
\end{figure}

The slow-roll parameters for the induced gravity model \eqref{eq:pot-induced} have the form
{\small
\begin{equation}
\epsilon_V = \frac{32 M_P^4 \left(\alpha  \lambda +\xi ^2\right) e^{\frac{4 \zeta  \sqrt{\alpha  \lambda +\xi ^2}}{\sqrt{\xi } M_P}} \left(2 \alpha  \lambda  M_P^2 e^{\frac{2 \zeta  \sqrt{\alpha  \lambda +\xi ^2}}{\sqrt{\xi } M_P}}+\xi  e^{\frac{4 \zeta  \sqrt{\alpha  \lambda +\xi ^2}}{\sqrt{\xi } M_P}}-\alpha  \lambda  \xi  M_P^4\right)^2}{\xi  \left(-2 \xi  M_P^2 e^{\frac{2 \zeta  \sqrt{\alpha  \lambda +\xi ^2}}{\sqrt{\xi } M_P}}+e^{\frac{4 \zeta  \sqrt{\alpha  \lambda +\xi ^2}}{\sqrt{\xi } M_P}}-\alpha  \lambda  M_P^4\right)^2 \left(e^{\frac{4 \zeta  \sqrt{\alpha  \lambda +\xi ^2}}{\sqrt{\xi } M_P}}+\alpha  \lambda  M_P^4\right)^2}\ ,
\end{equation} 
\begin{eqnarray}
\eta_V &=&     \frac{16 M_P^2 \left(\alpha  \lambda +\xi ^2\right)}{\xi } \left(\frac{6 \alpha ^2 \lambda ^2 M_P^6}{\left(e^{\frac{4 \zeta  \sqrt{\alpha  \lambda +\xi ^2}}{\sqrt{\xi } M_P}}+\alpha  \lambda  M_P^4\right)^2}+\frac{2 M_P^4 \left(\alpha  \lambda +\xi ^2\right) \left(2 \xi  e^{\frac{2 \zeta  \sqrt{\alpha  \lambda +\xi ^2}}{\sqrt{\xi } M_P}}+\alpha  \lambda  M_P^2\right)}{\left(-2 \xi  M_P^2 e^{\frac{2 \zeta  \sqrt{\alpha  \lambda +\xi ^2}}{\sqrt{\xi } M_P}}+e^{\frac{4 \zeta  \sqrt{\alpha  \lambda +\xi ^2}}{\sqrt{\xi } M_P}}-\alpha  \lambda  M_P^4\right)^2}       \right. \nonumber \\
&& \qquad \qquad \qquad \qquad \left. -\frac{6 \alpha  \lambda  M_P^2}{e^{\frac{4 \zeta  \sqrt{\alpha  \lambda +\xi ^2}}{\sqrt{\xi } M_P}}+\alpha  \lambda  M_P^4}+\frac{2 M_P^2 \left(\alpha  \lambda +\xi ^2\right)-\xi  e^{\frac{2 \zeta  \sqrt{\alpha  \lambda +\xi ^2}}{\sqrt{\xi } M_P}}}{-2 \xi  M_P^2 e^{\frac{2 \zeta  \sqrt{\alpha  \lambda +\xi ^2}}{\sqrt{\xi } M_P}}+e^{\frac{4 \zeta  \sqrt{\alpha  \lambda +\xi ^2}}{\sqrt{\xi } M_P}}-\alpha  \lambda  M_P^4}\right)\ .
\end{eqnarray} 
}
In Fig.~\ref{fig:r_ns-induced} we plot the predictions of the model in the $n_s-r$ plane for various values of the parameter $\alpha$, with fixed $\lambda = 0.01$, $\xi = 0.1$ and for $N=50-60$ e-folds. The dashed and solid lines represent the allowed $1\sigma$ and $2\sigma$ range of $n_s$,  set by the Planck 2018 collaboration~\cite{Aghanim2018, Akrami2018}. We observe that as $\alpha$ becomes smaller, we recover the predictions of the model without the $R^2$ term. Again, as $\alpha$ becomes larger the value of $r$ becomes smaller since the $R^2$ term dominates.

\subsection{The nonminimal Higgs model}
\label{sec:SR:Higgs}

The slow-roll parameters for the nonminimal Higgs model \eqref{eq:pot-induced} have the form 
{\small
\begin{equation}
\epsilon_V = \frac{8 \xi ^3 \left(\alpha  \lambda +\xi^2\right) \coth ^2\left(\frac{\zeta  \sqrt{\frac{\alpha  \lambda }{\xi }+\xi }}{M_P}\right) \text{csch}^4\left(\frac{\zeta  \sqrt{\frac{\alpha  \lambda }{\xi }+\xi }}{M_P}\right)}{\left(2 \xi ^2 \text{csch}^2\left(\frac{\zeta  \sqrt{\frac{\alpha  \lambda }{\xi }+\xi }}{M_P}\right)+M_P^2 \left(\xi ^2-\alpha  \lambda \right)\right)^2}\ ,
\end{equation}
\begin{equation}
\eta_V = \frac{4 \xi  \left(\alpha  \lambda +\xi^2\right) \text{csch}^4\left(\frac{\zeta  \sqrt{\frac{\alpha  \lambda }{\xi }+\xi }}{M_P}\right) \left(M_P^2 \left(\alpha  \lambda -\xi^2\right) \left(\cosh \left(\frac{2 \zeta  \sqrt{\frac{\alpha  \lambda }{\xi }+\xi }}{M_P}\right)+2\right)+2 \xi ^2 \left(\coth ^2\left(\frac{\zeta  \sqrt{\frac{\alpha  \lambda }{\xi }+\xi }}{M_P}\right)+1\right)\right)}{\left(2 \xi ^2 \text{csch}^2\left(\frac{\zeta  \sqrt{\frac{\alpha  \lambda }{\xi }+\xi }}{M_P}\right)+M_P^2 \left(\xi ^2-\alpha  \lambda \right)\right)^2}\ .
\end{equation}
}

\begin{figure}[H]
\includegraphics[width=.5\textwidth]{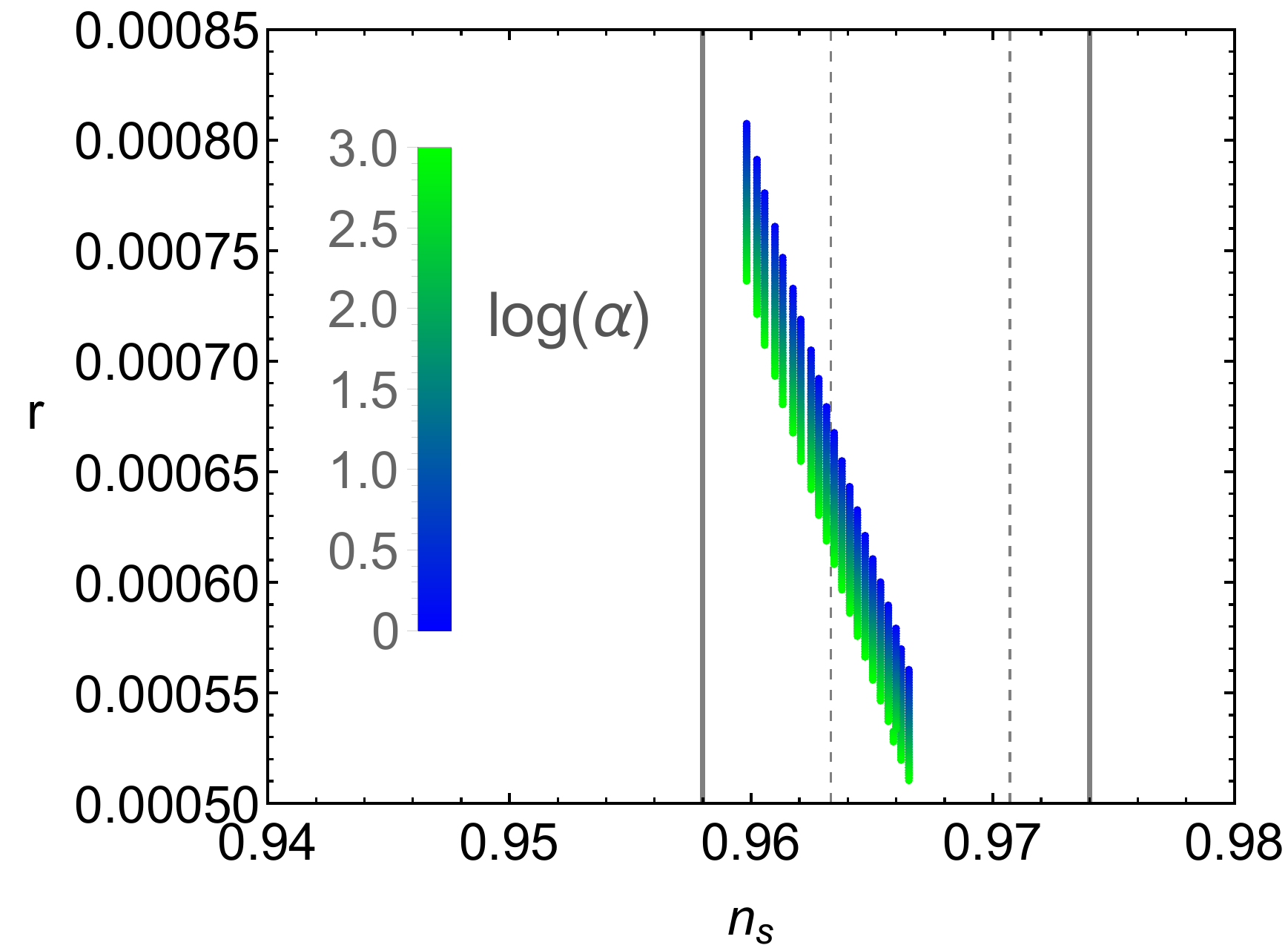}
\includegraphics[width=.5\textwidth]{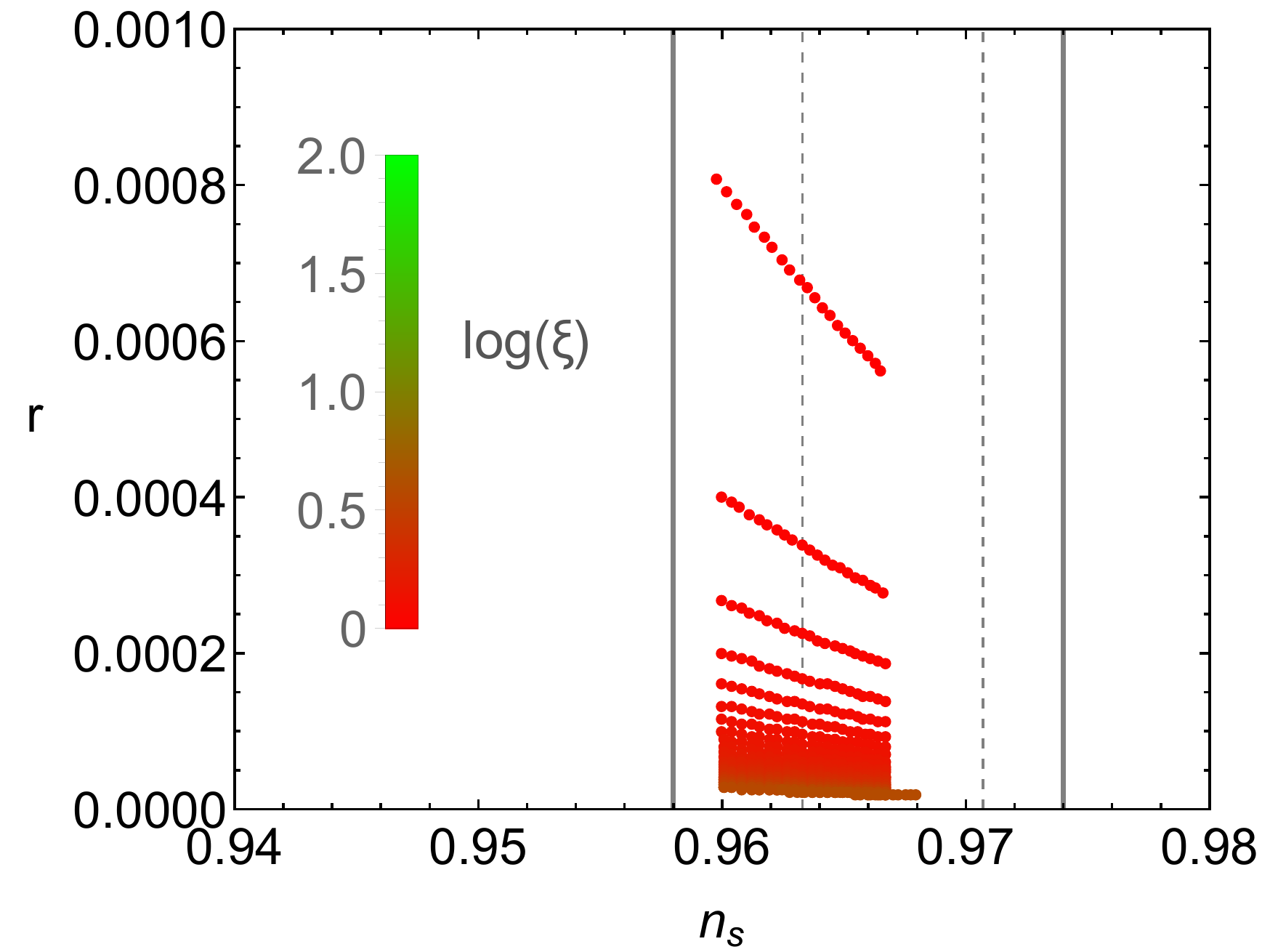}
\caption{The predictions for the inflationary observables in the $n_S-r$ plane for the nonminimal Higgs model with fixed $\lambda = 10^{-4}$ and for $N=50-60$ e-folds. On the left plot we fix $\xi = 1$ and vary the $R^2$ term parameter between $\alpha = 1-1000$. On the right plot we fix $\alpha=0.1$ and vary the nonminimal coupling between $\xi=0.001-1$.}
\label{fig:r_ns-higgs}
\end{figure}

In Fig.~\ref{fig:r_ns-higgs} we plot the predictions of the model in the $n_s-r$ plane with fixed $\lambda = 10^{-4}$, $\xi=1$ and for $N=50-60$ e-folds. We vary the $R^2$ term parameter between $\alpha = 1-1000$. Again, as $\alpha$ becomes larger the tensor-to-scalar ratio becomes smaller.

\section{Conclusions}
Among the few viable inflationary models, the Starobinsky model, in light of presently existing CMB data, has received particular attention. Nevertheless, in contrast to its standard metric formulation, within the Palatini formalism it does not provide us with a model for inflation, due to the absence of a propagating inflaton. In the present article we have considered the Starobinsky model in the framework of the Palatini formalism coupled to matter in the form of a scalar field, in general coupled nonminimally to gravity. The resulting Einstein frame theory contains non-trivial alterations in the scalar interactions that can drastically modify the inflationary behavior of the models considered. 

We have analyzed slow-roll inflation in a number of models and found that even models excluded within the standard formulation can be rendered viable due to the presence of the $R^2$ term in the Palatini formalism. Specifically, we have examined the model of a minimally coupled scalar with just a quadratic potential, a quasi-scale invariant Coleman-Weinberg model with a non-minimally coupled scalar, an induced gravity model and the nonminimally coupled Higgs model. 

We find that the minimally coupled quadratic model can provide a viable model of inflation. Subplanckian values of the model mass parameter ($m=0.1$) and not too-large values of the Starobinsky parameter ($\alpha\in[0.01,\,10]$) give $n_s$ and $r$ within the allowed region for a number of $e$-folds $50-60$. This is in contrast to the metric formulation of this model in the absence of the Starobinsky term which leads to a unacceptably large $r$. Next, we have analyzed the classically scale-invariant Coleman-Weinberg model and find it to be also a viable inflationary model. We find $n_s$ and $r$ within the acceptable region for a large range of $\alpha$ values and reasonably small $\xi\sim 0.1$, for which the model without the $R^2$ term in the metric formalism reaches the linear inflation limit. The induced gravity model of a nonminimally coupled scalar field with a quartic potential in which the Planck scale is generated by the scalar field VEV, has also been analyzed. We find that the predicted $n_s$ and $r$ values are within the acceptable region for values of the Starobinsky parameter $\alpha\in[0.1,\,10]$ and for reasonably small values of the couplings $\xi\sim 0.1$, $\lambda\sim 0.01$. Lastly, we have analyzed the non-minimally coupled Higgs model and found $n_s$ and $r$ values within the acceptable region for small $\xi\in[1,\,100]$ and $\lambda\sim\,10^{-4}$, $\alpha\sim1$. 

In conclusion, all the analyzed models can provide viable models of inflation, with the main effect of the $R^2$ term being that it lowers the value of $r$. The Palatini formalism has the advantage of providing single-field inflation, while in the metric formalism with an $R^2$ term we would need to consider multifield analysis. We expect more models that lie outside of the current allowed $r-n_s$ parameter space to follow the trend described above.

\section*{Acknowledgements}
This work was funded in part by the ``Institute Lagrange de Paris". I.A. is funded in part by the Swiss National Science Foundation and in part by a  CNRS PICS grant. A.K. acknowledges support from the Operational Program ``Human Resources Development, Education and Lifelong Learning" which is co-financed by the European Union (European Social Fund) and Greek national funds. A.L. acknowledges financial support by the Hellenic State Scholarship Foundation (IKY) through the action MIS 5000432. K.T. wishes to thank LPTHE for hospitality. 

\vspace{0.5cm}
{\centerline{\textbf{Note Added}}}
\noindent
While this article was prepared for submission, a related article by V.-E. Enckell et al.~appeared~\cite{Enckell2018a}, where the authors studied the effect of the $R^2$ term in the Palatini formalism.

\bibliography{References}{}
\bibliographystyle{utphys}

\end{document}